\documentclass[twocolumn,showpacs]{revtex4}

\usepackage{graphicx}
\usepackage{dcolumn}
\usepackage{bm}
\usepackage{hyperref}

\begin{document}

\title{Long term black hole evolution with the BSSN system 
by pseudospectral methods}

\author{Wolfgang Tichy}
\affiliation{Department of Physics, Florida Atlantic University,
             Boca Raton, FL  33431}




%
\newcommand\be{\begin{equation}}
\newcommand\ba{\begin{eqnarray}}

\newcommand\ee{\end{equation}}
\newcommand\ea{\end{eqnarray}}
\newcommand\p{{\partial}}
\newcommand\remove{{{\bf{THIS FIG. OR EQS. COULD BE REMOVED}}}}
%

\begin{abstract}

We present long term evolutions of a single black hole
of mass $M$ with the BSSN system using pseudospectral methods.
For our simulations we use the SGRID code
where the BSSN system is implemented in its standard
second order in space form. Previously we found 
that such simulations are quite unstable. The main
goal of this paper is to present two improvements
which now allow us to evolve for longer times.
The first improvement is related to the boundary conditions
at the excised black hole interior. We now use a gauge
condition that ensures that all modes are going into the black
hole, so that no boundary conditions are needed at the excision
surface. The second more significant improvement has to
do with our particular numerical method and involves
filters based on projecting the double Fourier expansions 
used for the angular dependence onto Spherical Harmonics.
With these two improvements it is now easily possible
to evolve for several thousand $M$. The only remaining
limitation seems to be the radiative outer boundary conditions
used here. Yet this problem can be ameliorated
by pushing out the location of the outer boundary, which leads
to even longer run-times.

\end{abstract}

\maketitle

\section{Introduction}

Currently several gravitational wave detectors such as 
LIGO~\cite{LIGO_web} or GEO600~\cite{GEO_web} are 
already operating, while several others are in the
planning or construction phase~\cite{Schutz99}. One of the most promising
sources for these detectors are the inspirals and mergers of binary black
holes. In order to make predictions about the final phase of such
inspirals and mergers, fully non-linear numerical simulations of the Einstein
Equations are required.
In order to numerically evolve the Einstein equations, at least
two ingredients are necessary. First we need a specific formulation
of the evolution equations. And second, a particular numerical
method such as finite differencing or a spectral method
is needed to implement these equations on a computer.
Currently there are two systems that have been used successfully
to evolve binary black hole systems. There is the generalized harmonic
system~\cite{Pretorius:2004jg}, which in its original second order
in space form has been only used in a finite
differencing code~\cite{Pretorius:2004jg,Pretorius:2005gq,
Pretorius:2006tp,Pretorius:2007jn,Palenzuela:2009yr}.
A first order version
of this system has also been successfully evolved using a spectral
method~\cite{Lindblom:2005qh,Scheel:2006gg,Boyle:2007ft,Scheel:2008rj,
Pazos:2009vb,Szilagyi:2009qz,Chu:2009md}.
The second and by far the most common system is
the BSSN system~\cite{Baumgarte:1998te}. This system
is second order in space. Over the last few years, many successful
binary black hole 
evolutions~\cite{Campanelli:2005dd,Baker:2005vv,
Baker:2006yw,Baker:2006vn,
Campanelli:2006gf,Campanelli:2006uy,Campanelli:2006fg,Campanelli:2006fy,
Gonzalez:2006md,Sperhake:2006cy,Campanelli:2007cga,Campanelli:2007ea,
Gonzalez:2007hi,Brugmann:2007zj,Herrmann:2007ex,Hinder:2007qu,
Koppitz:2007ev,Marronetti:2007ya,Marronetti:2007wz,
Pollney:2007ss,Rezzolla:2007xa,Rezzolla:2007rd,Rezzolla:2007rz,
Sperhake:2007gu,Tichy:2007hk,Dain:2008ck,Brugmann:2008zz,
Healy:2008js,Hinder:2008kv,Lousto:2008dn,Tichy:2008du,
Washik:2008jr,Bode:2009fq,Nakano:2009wk,Aylott:2009ya,Aylott:2009tn,
Pollney:2009yz,Reisswig:2009vc}
have been performed with this system.
However, long term simulations with BSSN have so far only 
been performed with finite differencing techniques. 
It seems thus natural to ask whether the BSSN system in its
standard second order form can also 
be implemented using a spectral method.
This question has been partially answered in
Ref.~\cite{Tichy:2006qn} (henceforth paper 1) 
where we have found that a single black hole can in principle
be evolved using BSSN together with a spectral method.
However, we found that our simulations fail after a short time,
usually after about $100M$, where $M$ is the mass of the black hole.
This result has been obtained by using the same gauge conditions
and outer boundary conditions for the spectral method as in 
successful finite difference implementations.
In this paper we describe two new ingredients which allow
us to evolve for much longer times with our spectral method.
The first one is a gauge condition that ensures that
no modes are leaving the black hole at the horizon, so that
no boundary condition is needed at the horizon.
The second ingredient is a particular spectral filter that
is based on projecting the double Fourier expansions
(used for the angular dependence of all fields) onto Spherical Harmonics.
This filter is very useful in removing unphysical modes
in our evolved fields and is the main reason for the
observed increased run-times.

Throughout we will use units where $G=c=1$.
The paper is organized as follows. 
In Sec.~\ref{methods} we describe briefly how a
single black hole can be evolved with the 
SGRID code~\cite{Tichy:2006qn,Tichy:2009yr}
and we introduce the two new ingredients
that lead to improved run-time.
Sec.~\ref{results} presents the results of several simulations using
these ingredients.
In Sec.~\ref{torustests} we show results from two code tests in a 3-torus.
We conclude with a discussion of our results in Sec.~\ref{discussion}.

\section{BSSN and the SGRID code}
\label{methods}

In this section we describe how a single black hole can
be evolved using the SGRID code. We first discuss the BSSN system
and then the spectral method we are using.

\subsection{A single black hole and the BSSN system}

As initial data we use a Schwarzschild black hole of mass $M$
in Kerr-Schild coordinates.
Thus initially the 3-metric and extrinsic curvature are given by
\begin{eqnarray}
g_{ij} &=& \delta_{ij} + 2 H l_i l_j \\
K_{ij} &=& \alpha [ l_{i} H_{,j} + l_{j} H_{,i} + H l_{i,j} + H l_{j,i} \nonumber \\
       & &+ 2 H^2 (l_{i} l_{k} l_{j,k} + l_{j} l_{k} l_{i,k}) 
          + 2 H l_{i} l_{j} l_{k} H_{,k} ] ,
\end{eqnarray}
where $H = M/r$ and $l^i = x^i/r$. The initial lapse and shift are
\begin{eqnarray}
\label{init_lapse}
\alpha = 1/\sqrt{1 + 2 H}, \\
\label{init_shift}
\beta^i = 2 l^i H/(1 + 2 H) .
\end{eqnarray}

The BSSN formulation introduces
a conformal metric $\tilde{\gamma}_{ij}$ (with determinant
one) and conformal factor $e^{4\phi}$, so that
$g_{ij} = e^{4\phi} \tilde{\gamma}_{ij}$.
It also introduce the additional variable 
$\tilde{\Gamma}^i
= \tilde{\gamma}^{ij} \tilde{\gamma}^{kl} \tilde{\gamma}_{jk,l}$.
The extrinsic curvature is split into its trace $K$
and tracefree part $\tilde{A}_{ij}$ using
$K_{ij} = e^{4\phi}
         \left( \tilde{A}_{ij} + K \tilde{\gamma}_{ij}/3 \right)$.
The BSSN evolution equations are then
\begin{eqnarray}
\label{dgdt}
\partial_t \tilde{\gamma}_{ij} 
&=& - 2 \alpha \tilde{A}_{ij} + \pounds_{\beta}\tilde{\gamma}_{ij} \\
\partial_t \phi 
&=& \frac{1}{6} \left( -\alpha K + D_i \beta^i \right) \\
\label{dGdt}
\partial_t \tilde{\Gamma}^i
&=& 
    - 2 \alpha \left(  \frac{2}{3} \tilde{\gamma}^{ij} D_j K 
              - 6 \tilde{A}^{ij} D_j \phi
              - \tilde{\Gamma}^{i}_{jk} \tilde{A}^{jk} \right) \nonumber \\
& & -2 \tilde{A}^{ij} D_j \alpha 
    -\frac{4}{3} ( \tilde{\Gamma}^i
       -\tilde{\gamma}^{jk} \tilde{\Gamma}^{i}_{jk}) \beta^l_{,l} 
    + \tilde{\gamma}^{jk} \beta^{i}_{,jk} \nonumber \\
& & + \frac{1}{3} \tilde{\gamma}^{ij} \beta^k_{,kj}
    - \tilde{\Gamma}^{j} \beta^i_{,j} 
    + \frac{2}{3} \tilde{\Gamma}^{i} \beta^k_{,k} 
    + \beta^j \tilde{\Gamma}^{i}_{,j} \\
\label{dAdt}
\partial_t \tilde{A}_{ij}
&=& e^{-4\phi} 
    \left[ - D_i D_j \alpha 
           + \alpha \left(  \tilde{R}_{ij} 
                          + R^{\phi}_{ij}\right) \right]^{TF} \nonumber \\
& & + \alpha (K \tilde{A}_{ij} - 2 \tilde{A}_{ik}\tilde{A}^{k}_{j}) 
    + \pounds_{\beta}\tilde{A}_{ij} \\
\label{dKdt}
\partial_t K
&=& - D^i D_i \alpha 
    + \alpha \left( \tilde{A}^{ij}\tilde{A}_{ij}+ \frac{1}{3} K^2 \right) 
    + \pounds_{\beta}K .
\end{eqnarray}
Here the superscript $TF$ in Eq.~(\ref{dAdt})
denotes the trace free part and $D_i$ is the derivative 
operator compatible with the $3$-metric. 
In order to ensure that $\tilde{A}_{ij}$
remains traceless during our numerical evolution, we subtract
any trace due to numerical errors after each evolution step
from $\tilde{A}_{ij}$.

As in paper 1 we will keep the shift $\beta^i$ constant
during all our evolutions and choose some particular coordinate
condition to evolve the lapse $\alpha$.
Furthermore we employ spherical coordinates 
to evolve all fields inside a spherical shell that is bounded
by a inner radius $R_{in}$ located just inside
the horizon and the outer radius $R_{out}$ that can be chosen
freely. This means that we need boundary conditions at both
the excision radius $R_{in}$ and at the outer boundary $R_{out}$.
At $R_{out}$ we use the same simple radiative boundary conditions
that were also employed in paper 1, 
and which are used by virtually all other BSSN implementations.
The difference in the current work lies at the excision
boundary $R_{in}$. In paper 1 
we had simply assumed that all modes at the excision radius $R_{in}$
are going into the hole (i.e. are leaving the numerical
domain), which implies that no boundary condition should be imposed
at $R_{in}$. However, as already shown by Bona et al.~\cite{Bona94b},
the gauge conditions of the ``1+log'' type used in paper 1
introduce faster than light gauge speeds so that one can
expect gauge modes to leave the hole.
By analyzing the characteristic modes of a
first order in space extension of the BSSN system 
Beyer and Sarbach~\cite{Beyer:2004sv} have explicitly
shown that the assumption that all modes are going into the hole
is only true for particular
gauge conditions and that in fact certain gauge modes
do leave the black hole for the gauges investigated in
paper 1. This result still holds for the second order in space
BSSN system used here. The reason is as follows:
The extension in~\cite{Beyer:2004sv} consists of merely 
introducing the additional derivative variables
$d_k = 12 \partial_k \phi$,
$d_{kij} = \partial_k \tilde{\gamma}_{ij}$,
$ A_k = (\partial_k \alpha)/\alpha $
\footnote{Any linear combination of these additional variables
would lead to the same mode speeds.}
and their evolution equations without adding any constraints.
Hence the mode speeds of this first order system are the
same as one would obtain from the formalism
of Gundlach and Martin-Garcia~\cite{Gundlach:2004ri,Gundlach:2004jp}
for second order systems, in which one would express the corresponding
modes in terms of longitudinal derivatives of 
$\phi$, $\tilde{\gamma}_{ij}$ and $\alpha$.

Thus not imposing any boundary conditions
at the excision radius while using the gauges in paper 1
will cause instabilities and is one
of the reasons why the runs in paper 1 failed
so early on. To avoid this problem we now use 
\begin{equation}
\label{lapse_evo}
\partial_t \alpha = \beta^i \partial_i \alpha - \alpha^2 K  + C_0 
\end{equation}
as the evolution equation for the lapse. Here the constant 
\begin{equation}
C_0 =  \alpha(t=0)^2 K(t=0) - \beta^i(t=0) \partial_i \alpha(t=0)
\end{equation}
is computed from the initial values of $\alpha$, $\beta^i$ and $K$
and is chosen such that $\partial_t \alpha = 0$ initially.
Using the equations for the characteristic speeds
given in~\cite{Beyer:2004sv} it is clear that 
all modes (including the gauge modes) of the first order BSSN version
are going into the black hole inside the horizon when the
lapse evolves according to Eq.~(\ref{lapse_evo}). 
Therefore it is now indeed justified to not impose
any boundary conditions at $R_{in}$, which is precisely what
we will do here. This observation
carries over to the second order in space version of the BSSN
system we are using here.
However, this new gauge condition alone does not lead
to a big improvement in the run-time of our code.
We also need the filter algorithm described in the next subsection.

\subsection{Numerical methods employed in the SGRID code}
\label{specfilter}

For the time integration of all evolved fields
we use a fourth order accurate Runge-Kutta time integrator.
In order to compute spatial derivatives
we apply essentially the same pseudospectral
collocation method as in paper 1
but this time we add a particular filter algorithm that
leads to big increases in run-time.
For a thorough discussion of collocation methods and other possible
spectral methods in a more general setting we refer the reader
to a recent review article by Grandcl\'{e}ment and 
Novak~\cite{Grandclement:lrr-2009-1}.

As in paper 1 we use standard spherical coordinates and
use Chebyshev polynomials in the radial direction and Fourier expansions
in both angles. The collocation points are then given by
\begin{eqnarray}
\label{r_i}
r_i      &=& \frac{R_{in}-R_{out}}{2}\cos\left(\frac{i \pi}{N_{r} -
1}\right)
            +\frac{R_{in}+R_{out}}{2} \\
\label{theta_j}
\theta_j &=& \pi (2j+1)/N_{\theta} \\
\label{varphi_k}
\varphi_k   &=& 2\pi k/N_{\varphi},
\end{eqnarray}
where $0\leq i \leq N_{r} -1$, $0\leq j \leq N_{\theta} -1$,
$0\leq k \leq N_{\varphi} -1$ and $N_{r}$, $N_{\theta}$, $N_{\varphi}$
are the number of collocation points in each direction.
Notice that both angles run from 0 to $2\pi$, which is necessary
to ensure the periodicity in both angles needed for
Fourier expansions.
In paper 1 we have found that this double
covering causes no problems when we evolve scalar fields.
Furthermore, simply removing the double covering does not improve
the run-time for a simulation with BSSN.

In order to avoid problems near the coordinate singularities at the poles
we evolve the Cartesian components of all BSSN fields.
The spatial components of the evolved fields thus are
$\phi$, $\tilde{\gamma}_{ij}$, $\tilde{\Gamma}^{i}$,
$\tilde{A}_{ij}$, and $K$.
We can expand these fields in terms of spherical harmonics. 
For a spherically symmetric hole the modes contained in these
fields are 
$l=0$ for $\phi$ and $K$,
$l=0$ and $l=1$ for $\tilde{\Gamma}^{i}$,
$l=0$, $l=1$ and $l=2$ for $\tilde{\gamma}_{ij}$ and $\tilde{A}_{ij}$.
This shows that scalar, vector and tensor like fields
contain a different number of $l$-modes.
However, our double Fourier expansions of both angles do not make this
distinction, and in addition they even allow for fields
that could not be represented by Spherical Harmonics
(because of the double covering).
This motivates the following spectral filter algorithm,
which we have found to be crucial to extend the run-time of our simulations.
From the representation of each field $u$ by its values at the collocation
points we compute expansion coefficients in terms of
discrete Spherical Harmonics. I.e. for each radial grid point
we compute~\cite{Healy2003}
\begin{equation}
\label{DSHT}
c(r_i)_{lm} = \frac{\sqrt{2\pi}}{2B}
\sum_{j=0}^{2B-1} w_j P_l^m (\cos \theta_j)
\sum_{k=0}^{N_{\varphi}-1} e^{im\varphi_k} u(r_i, \theta_j, \varphi_k)
\end{equation}
where the $P_l^m(x)$ are Associated Legendre polynomials, their
weights are
\begin{eqnarray} 
w_j &=& \frac{2}{B} \sin\left(\frac{(2j+1)\pi}{4B}\right) \times\nonumber \\
    & & \sum_{k=0}^{B-1} \frac{1}{2k+1}
                       \sin\left((2j+1)(2k+1)\frac{\pi}{4B}\right) ,
\end{eqnarray}
$B=N_{\theta}/4$ and we choose $N_{\theta}$ to be
a multiple of 4.
The sum over $k$ in Eq.~(\ref{DSHT}) is a simple Fourier transform,
while the sum over $j$ is computed using routines from the S2kit
package~\cite{S2kit_web,Healy2003}. 
The maximum $l$
in this expansion is $l_{max} = B-1 = N_{\theta}/4-1$, and as usual
we set $c(r_i)_{lm}=0$ if $m > l_{max}$.
Note that the sum over $k$ in Eq.~(\ref{DSHT}) only runs up
to $2B-1 = N_{\theta}/2-1$, i.e. the only field values taken
into account come from points with $0 \leq \theta \leq \pi$.
From the coefficients $c(r_i)_{lm}$ we can then recompute
$u(r_i, \theta_j, \varphi_k)$ via the inverse discrete Spherical 
Harmonic transform.

In terms of a double Fourier representation
an expansion up to $l_{max}$ corresponds to 
$2l_{max}+1$ Fourier modes for both the $\theta$- and $\varphi$-directions.
I.e. after recomputing the $u(r_i, \theta_j, \varphi_k)$ 
via the inverse transform only $2l_{max}+1=N_{\theta}/2-1$
Fourier modes remain. Since originally we had $N_{\theta}$
Fourier modes, this operation filters out about half of all
Fourier modes for the $\theta$-direction.
Similarly it filters out all modes above $k=2l_{max}=N_{\theta}/2-2$
for the $\varphi$-direction.
For our simulations we choose $N_{\varphi}=3N_{\theta}/4$. In that
case the above projection onto Spherical Harmonics filters
the upper third of the Fourier modes for the $\varphi$-direction.
We use this procedure to filter
all tensor like fields (i.e. $\tilde{\gamma}_{ij}$ and $\tilde{A}_{ij}$).
For vector like fields ($\tilde{\Gamma}^{i}$ and $\beta^i$)
we additionally set $c(r_i)_{l_{max}m}=0$ so that we filter one 
more $l$-mode. For scalars like fields ($\phi$, $K$ and $\alpha$).
we set $c(r_i)_{l_{max}m}=c(r_i)_{l_{max}-1\ m}=0$
so that we filter yet one more $l$-mode.
Notice that this filter algorithm also removes the double covering.
We apply it after each evolution step. It removes any unphysical
modes (i.e. the ones which cannot be represented by an
expansion of Spherical Harmonics). It also removes high frequency
modes that have been contaminated by aliasing
due to the non-linear terms in the BSSN system.
Since the submission of this paper we have also successfully used
this filter in scalar field evolutions~\cite{Vega:2009qb}.

When we apply the filter algorithm for the angular directions
detailed above our code runs noticeably longer. 
Since the BSSN system is non-linear, aliasing
also to occurs in the radial direction. 
We can further improve the run-time by applying
the standard Orszag 2/3 
rule~\cite{Orszag2o3rule,Boyd00} in the radial direction:
We first compute the coefficients $a_i$ of the Chebyshev expansion in the 
radial direction for each field and then set $a_i=0$ for all $i\geq 2N_r/3$,
i.e. we filter out the top 1/3 of the Chebyshev modes.
From the so filtered $a_i$ we recompute the values of all fields
at the collocation points.
When we apply both the radial and especially the angular filter
algorithms we get a greatly improved run-time.

\section{Evolving a black hole with the SGRID code}
\label{results}

%
%
%
%
%
%

In this section we show the results from several simulations
of a single black hole. The Kerr-Schild initial data together
with the choice of lapse and shift explained above
describe a static black hole, so that during evolution
all fields should remain constant. However due to numerical
errors this is not exactly the case. In fact we observe
that the norms of all fields grow in time until the code finally
fails. This growth is exponential as is typical for
the numerical instabilities in such problems.
Nevertheless, the time scale for this growth is now much longer
than in paper 1.

In all the simulation described in this paper the time step
in our Runge-Kutta time integrator is set to
\begin{equation}
\Delta t = \Delta r_{min}/2 ,
\end{equation}
where $\Delta r_{min}$ is the distance between the two closest
grid points in the radial direction. 
For the first set of simulations
we use a spherical shell that extends from $R_{in}=1.85M$ to
$R_{out}=480M$ as our numerical domain. 
The angular
resolution is chosen to be $(N_{\theta},N_{\varphi})=(16,12)$,
which corresponds to $l_{max}=3$ after applying our filters.
\begin{figure}
\includegraphics[scale=0.33,clip=true]{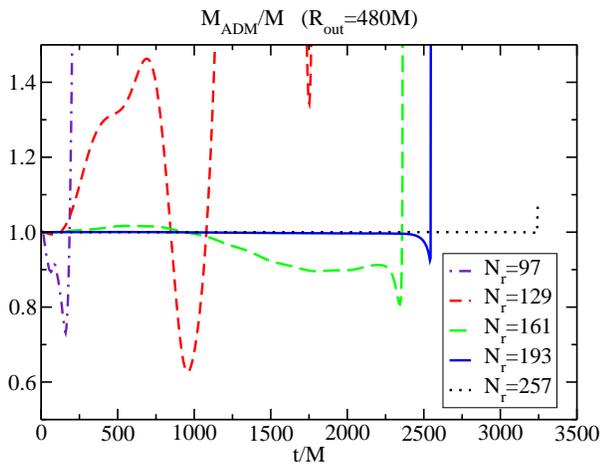}
\caption{\label{E_ADM_vs_res}
This plot shows the ADM mass $M_{ADM}$ for simulations
with different radial resolutions. In each case $R_{out}=480M$ and
$(N_{\theta},N_{\varphi})=(16,12)$.
The code performs much better for higher resolutions.
}
\end{figure}
\begin{figure}
\includegraphics[scale=0.33,clip=true]{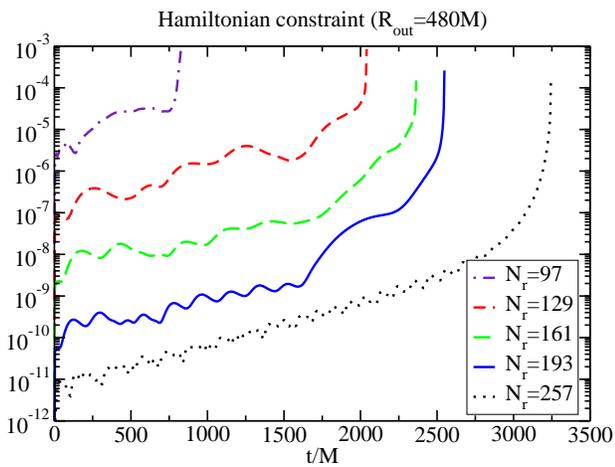}
\caption{\label{ham_vs_res}
This plot shows the $L^2$-norm of the Hamiltonian constraint violation
for different radial resolutions. In each case $R_{out}=480M$ and
$(N_{\theta},N_{\varphi})=(16,12)$. We observe geometric convergence 
with increasing resolution.
}
\end{figure}
Figures \ref{E_ADM_vs_res} and \ref{ham_vs_res}
show how the ADM mass $M_{ADM}$ and the Hamiltonian constraint
\begin{equation}
\label{ham_def}
H = R + K^2 - K^i_j K_i^j
\end{equation}
evolve in time for different radial resolutions
(given in terms of $N_r$). In all cases examined, our code fails
eventually. The time when it fails corresponds to the end of the lines
in Fig.~\ref{ham_vs_res}. However, the run-time in
each case is significantly longer than in paper 1.
Furthermore, we see from both figures that the code runs longer
for higher resolutions. For example for $N_r=257$ we get a run-time
of $3244M$, which is more than 30 times longer than in
paper 1. As is evident from Fig.~\ref{ham_vs_res}
our code exhibits the geometric convergence expected from
a spectral code. From Fig.~\ref{E_ADM_vs_res} we see that
the $M_{ADM}$ (which should in principle be always equal to $M$)
remains closer to $M$ for longer with increased resolution.

The simulations discussed so far were all for the same angular
resolution.
\begin{figure}
\includegraphics[scale=0.33,clip=true]{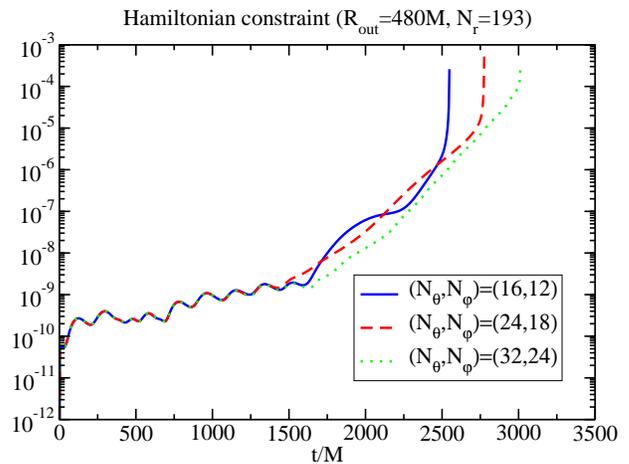}
\caption{\label{ham_vs_ang}
This plot shows the $L^2$-norm of the Hamiltonian constraint violation
for different angular resolutions. In each case $R_{out}=480M$ and
$N_r=193$. We see that the results do not depend much on
the angular resolution.
}
\end{figure}
However, as one can see from Fig.~\ref{ham_vs_ang}, our results
are largely independent of angular resolutions as one would
expect for a spherically symmetric problem. For this reason
we will revert to $(N_{\theta},N_{\varphi})=(16,12)$ from now on.

The reader may wonder why in fact our code fails at
late times. The reason lies with the radiative outer
boundary conditions we use when we evolve the BSSN system.
\begin{figure}
\includegraphics[scale=0.33,clip=true]{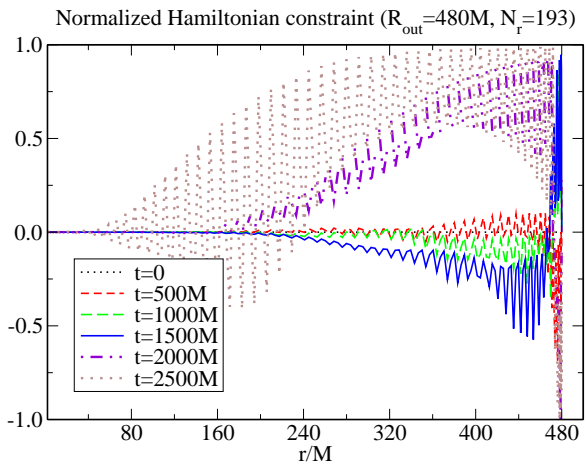}
\caption{\label{normham_vs_r}
This plot shows the normalized Hamiltonian along the radial
direction ($\theta=9\pi/16$, $\varphi=0$) for different times. The resolution is 
$(N_r,N_{\theta},N_{\varphi})=(193,16,12)$ and the
outer boundary is at $R_{out}=480M$.
We see how constraint violations enter through 
the outer boundary and in time contaminate the entire numerical domain.
}
\end{figure}
Figure~\ref{normham_vs_r} depicts the normalized Hamiltonian constraint,
which is obtained from the Hamiltonian constraint divided by
the magnitude of all terms that enter Eq.~(\ref{ham_def}).
Compared to $H$ this normalization magnifies errors near the outer
boundary, since far away from the black hole the individual terms
in Eq.~(\ref{ham_def}) are all small.
Figure~\ref{normham_vs_r} shows that Hamiltonian constraint violations
which are at first ($t=500M$) concentrated near the outer boundary
gradually enter the numerical domain, until the simulation fails
shortly after $t=2500M$.
In principle this problem should not come as a surprise.
The standard radiative outer boundary conditions
impose conditions on all evolved fields, which implies that
we are imposing conditions on all ingoing and outgoing modes.
However, we really should only impose conditions on the
ingoing modes.
In paper 1 we have shown that for a scalar field
system the radiative conditions are equivalent to imposing 
conditions only on incoming modes, because the conditions on the 
outgoing modes are equivalent to the evolution equations. 
For the BSSN system, however, there is no clear cut answer.
To our knowledge the well-posedness of the boundary problem for
the BSSN system has never been established and remains an open question.
The work by Beyer and Sarbach~\cite{Beyer:2004sv} mentioned above
comes closest to addressing our problem. They provide explicit
boundary conditions for the incoming modes of the BSSN system
for the case of a frozen shift that is tangential to the boundary.
These boundary conditions lead to a well posed initial-boundary value
formulation. One problem is that our shift is not tangential
to the outer boundary so that their proof does not strictly apply here.
Yet even if the proof could be extended to a radial shift, another
possibly more serious problem remains. Beyer and Sarbach provide
only seven boundary conditions because there are exactly seven
modes that enter through the outer 
boundary~\footnote{This fact also holds for a shift that is not
tangential.}. However, the radiative boundary conditions we use
here impose conditions on all evolved variables, which amounts to
18 conditions. Simple counting arguments make it seem unlikely
that our 18 conditions reduce to only 7 conditions on the incoming modes.
So if indeed there is a problem with well-posedness at the continuum
level, the real surprise is that radiative outer boundary conditions
work as well as they do for the BSSN system.
Recall that all the finite differencing codes
that have recently been used to perform long term simulations
of binary black holes~\cite{Campanelli:2005dd,Baker:2005vv,
Baker:2006yw,Baker:2006vn,
Campanelli:2006gf,Campanelli:2006uy,Campanelli:2006fg,Campanelli:2006fy,
Gonzalez:2006md,Sperhake:2006cy,Campanelli:2007cga,Campanelli:2007ea,
Gonzalez:2007hi,Brugmann:2007zj,Herrmann:2007ex,Hinder:2007qu,
Koppitz:2007ev,Marronetti:2007ya,Marronetti:2007wz,
Pollney:2007ss,Rezzolla:2007xa,Rezzolla:2007rd,Rezzolla:2007rz,
Sperhake:2007gu,Tichy:2007hk,Dain:2008ck,Brugmann:2008zz,
Healy:2008js,Hinder:2008kv,Lousto:2008dn,Tichy:2008du,
Washik:2008jr,Bode:2009fq,Nakano:2009wk,Aylott:2009ya,Aylott:2009tn}
also use these radiative outer boundary conditions.
These finite differencing codes are typically run for several thousand $M$
without failing. However, they also have problems at the
outer boundary in the sense that constraint violations are observed
to enter.
So we conclude that we need improved outer boundary conditions
if we want to evolve for even longer times, at least with a spectral
code.
Nevertheless, an evolution time of about $2500M$ is not all that
bad and may already be sufficient for many problems.
In addition, we have found a simple way to further increase the run-time.
\begin{figure}
\includegraphics[scale=0.33,clip=true]{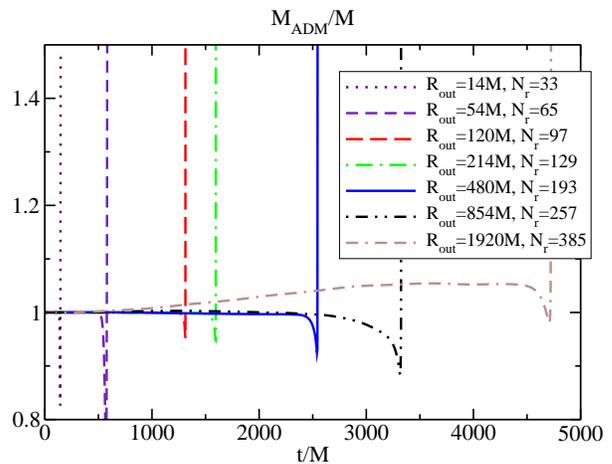}
\caption{\label{E_ADM_vs_Rout}
This plot shows the ADM mass for runs with different $R_{out}$.
In each case we use $(N_{\theta},N_{\varphi})=(16,12)$ and
adjust $N_r$ such that the resolution near the black hole
is the same for each run. The lines end when the run fails.
We observe that the run-time increases with increasing $R_{out}$.
}
\end{figure}
Figure~\ref{E_ADM_vs_Rout} shows the ADM mass $M_{ADM}$ for
runs with different $R_{out}$. The lines end when the code fails.
In all these runs we use the same
angular resolution and we adjust $N_r$ such that
the radial resolution is kept constant near the black hole horizon.
We see that the run-time is longer for larger $R_{out}$.
Thus we are able to increase the run-time as much as we need.
The only cost is the additional computer time needed to simulate
a larger domain. For example for $R_{out}=1920M$ we can evolve
for more than $4500M$.
Note that while the radial resolution is the same near the black hole
horizon for all runs shown in Fig.~\ref{E_ADM_vs_Rout}, the resolution
in the middle of the domain is less for the runs with larger $R_{out}$.
This fact is related to the uneven spread of collocation points for
the Chebyshev expansions used in the radial direction.
Thus effectively the runs with larger $R_{out}$ are less accurate.
We could have of course increased $N_r$ even further so that
the resolution in the middle of the domain would always be the same.
In that case the radial resolution near the black hole would increase
with $R_{out}$ and we would get even longer run-times, since
as evidenced by Fig.~\ref{ham_vs_res} the run-time only increases
with increased resolution.
\begin{figure}
\includegraphics[scale=0.33,clip=true]{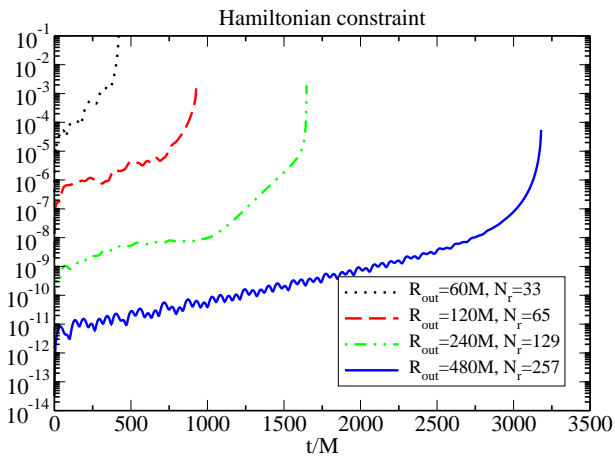}
\caption{\label{ham_vs_Rout2}
This plot shows the Hamiltonian constraint for runs with different $R_{out}$.
In each case we use $(N_{\theta},N_{\varphi})=(16,12)$ and
adjust $N_r$ such that the resolution in the middle of the domain
is the same for each run. The lines end when the run fails.
We observe that the run-time is proportional to $R_{out}$.
}
\end{figure}
Figure~\ref{ham_vs_Rout2} is the analog of Fig.~\ref{E_ADM_vs_Rout}
for the case where we increase $N_r$ proportional to $R_{out}$.
We see that the run-time is now approximately proportional to $R_{out}$.

\begin{figure}
\includegraphics[scale=0.33,clip=true]{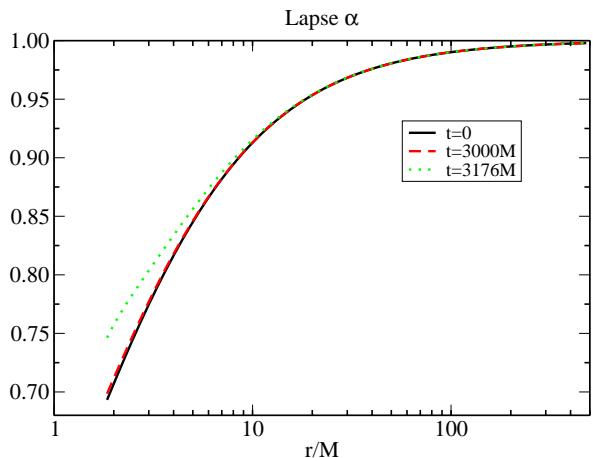}
\caption{\label{alpha_of_t}
This plot shows the lapse $\alpha$ along the radial direction
for the run with $R_{out}=480M$ and
$(N_r,N_{\theta},N_{\varphi})=(257,16,12)$ for three different times.
}
\end{figure}

In order to get a sense of how much the metric fields change during
our evolutions Fig.~\ref{alpha_of_t} depicts the lapse $\alpha$
at three different times for the longest run of Fig.~\ref{ham_vs_Rout2}.
From Fig.~\ref{alpha_of_t} we see that until $t=3000M$ the lapse 
does not change by much. Most of the change occurs just before the
run fails at $3176M$.

In order to study the angular dependence of the constraint
violations entering through the outer boundary we next show
snapshots of the normalized Hamiltonian constraint in the $xz$-plane.
\begin{figure}
\includegraphics[scale=0.7,clip=true]{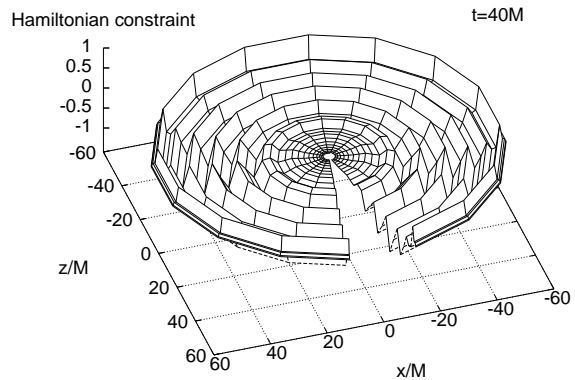}
\caption{\label{normham_vs_xz40}
The normalized Hamiltonian in the $xz$-plane at $t=40M$ for a run
using our spectral filters.
The resolution is $(N_r,N_{\theta},N_{\varphi})=(33,16,12)$ and the
outer boundary is at $R_{out}=60M$.
We see how constraint violations have entered through 
the outer boundary.
}
\end{figure}
\begin{figure}
\includegraphics[scale=0.7,clip=true]{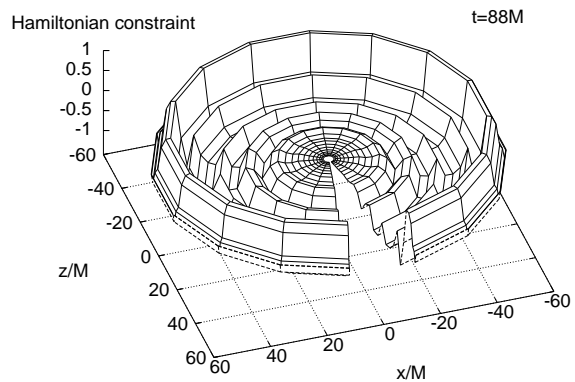}
\caption{\label{normham_vs_xz88}
Same as Fig.~\ref{normham_vs_xz40} but at $t=88M$.
}
\end{figure}
\begin{figure}
\includegraphics[scale=0.7,clip=true]{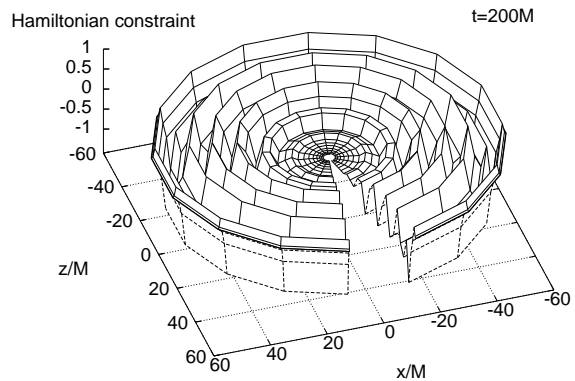}
\caption{\label{normham_vs_xz200}
Same as Fig.~\ref{normham_vs_xz40} but at $t=200M$.
}
\end{figure}
\begin{figure}
\includegraphics[scale=0.7,clip=true]{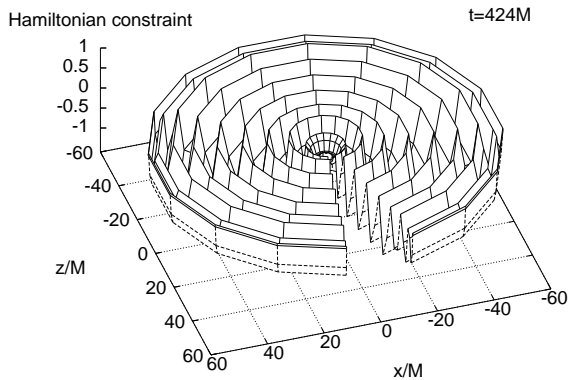}
\caption{\label{normham_vs_xz424}
Same as Fig.~\ref{normham_vs_xz40} but at $t=424M$ just before the run fails. 
}
\end{figure}
\begin{figure}
\includegraphics[scale=0.7,clip=true]{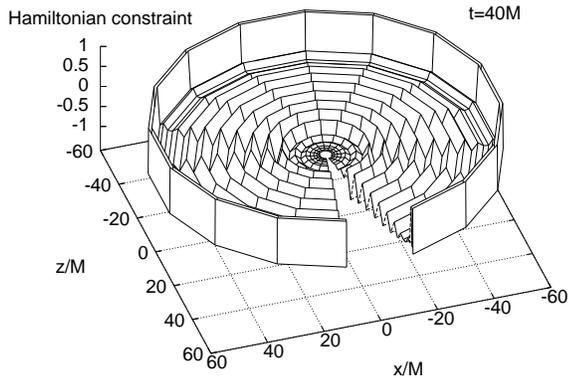}
\caption{\label{NoFilt_normham_vs_xz40}
Same as Fig.~\ref{normham_vs_xz40} but without filtering.
}
\end{figure}
\begin{figure}
\includegraphics[scale=0.7,clip=true]{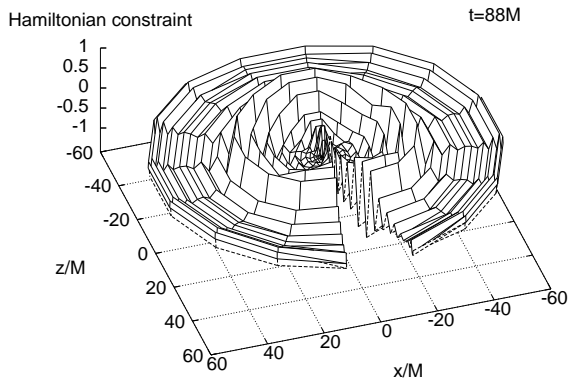}
\caption{\label{NoFilt_normham_vs_xz88}
Same as Fig.~\ref{NoFilt_normham_vs_xz40} but at $t=88M$ just before
the run fails. No filters were used. The result is no longer 
spherically symmetric.
}
\end{figure}
For visual clarity we depict results from a short low resolution
run with outer boundary at $R_{out}=60M$.
Figures~\ref{normham_vs_xz40}-\ref{normham_vs_xz424} show that
the constraint violations have no strong angular dependence
when our filters are active. Yet we see strong oscillations in
the radial direction. This particular simulation fails shortly after
$t=424M$. Figures~\ref{NoFilt_normham_vs_xz40}
and \ref{NoFilt_normham_vs_xz88} show how our results change if our
filters are switched off. The constraint violations are larger, the run
fails sooner (shortly after $t=88M$), and towards the end
our results deviate significantly from spherical symmetry.
Nevertheless the strongest oscillations are still in the radial direction.

\section{BSSN tests in a 3-torus}
\label{torustests}

As we have seen our black hole runs always fail at some point.
In order to strengthen the conjecture that this failure
is related to the radiative outer boundary conditions,
we have also performed two simple tests in a 3-torus, i.e. 
without boundary conditions. These tests were originally proposed
in~\cite{Alcubierre2003:mexico-I}. For the BSSN system both tests
have been performed before~\cite{Jansen:2003uh} using a finite
difference code~\cite{Bruegmann:2003aw,Bruegmann:2006at}. 
As we will see below the spectral SGRID code
is at least as stable as the finite differencing code.
Both tests are carried out in Cartesian
coordinates using Fourier expansions in $x$, $y$ and $z$-directions.
Also, both are effectively one-dimensional as we only use either 3 or 1
points in the $y$ and $z$-directions.

\subsection{Random perturbations on flat space}

In this test we evolve flat space in a 3-torus with added random
perturbations. The initial data are given by~\cite{Alcubierre2003:mexico-I} 
\begin{eqnarray}
g_{ij} &=& \delta_{ij} + \varepsilon_{ij} , \\
K_{ij} &=& \varepsilon_{ij}, \\
\alpha &=& 1 + \varepsilon_{ij}, \\
\beta^i &=& 0 ,
\end{eqnarray}
where $\varepsilon_{ij}$ is a random number with a 
probability distribution that is uniform in the 
interval $[-10^{-10}/(50/N_x)^2,+10^{-10}/(50/N_x)^2]$. 
Hence our initial data differ
slightly from the ones in~\cite{Jansen:2003uh} who add random noise 
only to $g_{ij}$.
The parameters for this test are:
\begin{itemize}
\item Points: $N_x$ is varied, $N_y = N_z = 3$
\item Resolution $\Delta x = 1/N_x$
\item Simulation domain: $x \in [0,1]$, \  
                         $(y,z) \in [0,3\Delta x]$
\item Gauge: 
  $\partial_t \alpha = -\alpha^2 \mbox{tr}K$, $\beta^i=0$
\end{itemize}
%
\begin{figure}
\includegraphics[scale=0.33,clip=true]{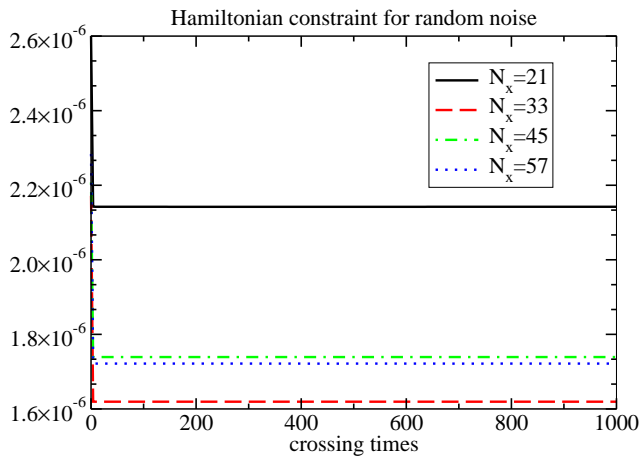}
\caption{\label{3dRandomNoise_ham}
The $L^2$-norm of the Hamiltonian constraint does not grow in
time if we evolve flat space with random noise in a 3-torus. 
As expected, the Hamiltonian constraint has approximately
the same value for all resolutions.
}
\end{figure}
Our results are shown in Fig.~\ref{3dRandomNoise_ham}.
As we can see our simulations are all stable. The Hamiltonian constraint
violation is constant. As expected the Hamiltonian constraint does
not converge to zero, since we have added random perturbations to flat
space, which do not satisfy the constraints. The amplitude of the
added random noise was chosen such that the Hamiltonian constraint
violation should be independent of the spatial resolution. 
From Fig.~\ref{3dRandomNoise_ham} we see that we indeed get 
roughly the same constraint violation for each run.
The same qualitative results for BSSN were obtained in~\cite{Jansen:2003uh} 
with a finite differencing code. We should also note that this
test is non-trivial. In~\cite{Jansen:2003uh} it was shown that
(at least in the finite differencing case) both the ADM~\cite{Arnowitt62}
and the AA~\cite{Alekseenko2002} system fail this test.

\subsection{Gauge wave}

In this test we look at the stability of the BSSN system in the non-linear
case. We consider flat space in a coordinate system where the metric 
is given by~\cite{Alcubierre2003:mexico-I} 
\begin{equation}
ds^2 = -(1+a)dt^2 + (1+a)dx^2 + dy^2 + dz^2,
\end{equation}
with $a = A\sin[2(x-t)/L]$.
We set our initial data using this metric at $t=0$.
Here $L$ is the size of the domain in the $x$-direction 
and $A$ is the amplitude of the wave.
Since this wave propagates along the $x$-axis and all derivatives are
zero in the $y$- and $z$-directions, the problem is one-dimensional. 
For our tests we use these parameters:
\begin{itemize}
\item Points: $N_x$ is varied, $N_y = N_z = 1$
\item One-dimensional simulation domain: $x \in [0,1]$
\item $A=0.01$ 
\item Gauge:
  $\partial_t \alpha = -\alpha^2 \mbox{tr}K$, $\beta^i=0$
\end{itemize}
\begin{figure}
\includegraphics[scale=0.33,clip=true]{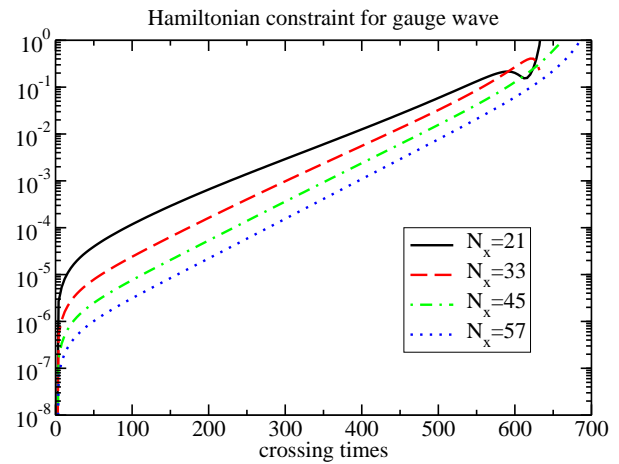}
\caption{\label{GaugeWave_ham}
The $L^2$-norm of the Hamiltonian constraint grows exponentially
for a gauge wave in a 3-torus. Convergence is lost after about 600
crossing times. The lines end when the runs fail. Nevertheless
the runs last about seven times longer than with a finite difference code.
}
\end{figure}
From Fig.~\ref{GaugeWave_ham} we see that our simulations are unstable.
We lose convergence after about 600 light crossing times. After that the runs
fail at around 700 crossing times. While this result is disappointing,
it is not surprising. When Jansen et al.~\cite{Jansen:2003uh} performed
the very same test with a finite differencing code, BSSN failed
even earlier at around 100 crossing times. So in fact the SGRID
code runs longer than a finite differencing code and the 
observed instability is likely not a result of our numerical method.
Furthermore, the gauge wave test is a member of a one-parameter family of
exponential harmonic gauge solutions~\cite{Babiuc:2005fr,Paschalidis:2007cp}.
Any numerical error can probe a member that is exponentially
growing. Thus generically we expect an exponential blow up for any
formulation with any code.

\subsection{Test results}

We should point out that for the two test results presented
above we have used the Orszag 2/3 rule in the $x$-direction
to filter out high frequency Fourier modes.
The motivation for this filter was the success of the filters 
in our black hole evolution. We have found, however, that this filter
does not help and that we obtain qualitatively the same results
without it.

As we have seen, the spectral implementation of BSSN performs
as well or better than a finite differencing implementation.
In the random noise test
the BSSN implementation in SGRID shows long term stability.
In the gauge wave test SGRID runs longer than a finite
differencing code, but fails after about 700 crossing times.
We note that our black hole simulations last only a few crossing times.
Thus when measured in crossing times BSSN in SGRID runs much longer
without boundary conditions than with radiative boundary conditions.
We interpret this fact as further evidence for our conjecture
that the radiative outer boundary conditions are the main reason
for the failure of our black hole evolutions.

\section{Discussion}
\label{discussion}

As we have seen long term evolutions of single black holes
with the BSSN system are possible with a spectral method.
One of the ingredients needed to achieve this goal
is a suitable gauge condition that is compatible with
imposing no boundary conditions at the excision
surface. The second and main ingredient is the spectral filter
described in Sec.~\ref{specfilter} above. This filter
is capable of removing unphysical modes
related to the double covering introduced by double
Fourier expansions. At the same time it
also removes high frequency modes that are contaminated
by aliasing due to non-linear terms in the BSSN system.
Simulations with this filter run significantly longer.
The only limitation seem to be problems
at the outer boundary coming from the simplistic radiative
conditions used there. This shows that further work 
along the lines of~\cite{Calabrese:2002xy}
is in principle needed
to determine better outer boundary conditions for the
BSSN system. There has been some 
work~\cite{Gundlach:2004ri,Gundlach:2004jp,Gundlach:2005ta}
on determining ingoing and outgoing
modes for the BSSN system in its standard second order in space form.
Also, a recent paper by N\'u\~nez and Sarbach~\cite{Nunez:2009wn}
presents interesting new boundary conditions that lead to a well posed 
initial-boundary value formulation for the case of linearized gravity.
However, we are not aware of any publication that 
provides explicit boundary conditions for BSSN which
we could directly implement in our code.
And deriving such boundary conditions is beyond the scope of this work.
But as we have seen, the problem of not having good outer boundary conditions
can also be circumvented by pushing out the location of the outer boundary.
In addition, a run-time of several thousand $M$ may be
completely sufficient for many applications.
For example, the many simulations 
we recently performed with our finite differencing code
to determine the final mass and spin of black hole 
mergers~\cite{Tichy:2008du} all just needed a run-time between
$400M$ and $1000M$.
So it seems that the main roadblock for long term simulations
of the BSSN system with spectral methods has been removed
by the filter algorithms introduced in Sec.~\ref{specfilter}.

Nevertheless, our work provides a strong motivation
for finding better outer boundary conditions for the BSSN system.
We have observed that radiative outer boundary conditions cause
constraint violations to enter the numerical domain. This happens
for both spectral and finite differencing codes.
Furthermore, BSSN in our spectral code runs for only a few
light crossing times when we evolve with radiative outer
boundary conditions, while it runs for hundreds of crossing times
without these boundary conditions. So far this instability has not
been observed with finite differencing codes. One reason for
this difference may be that finite differencing codes are local
since their stencils cover only a few points. Spectral
codes, on the other hand, are global in the sense that their
stencils cover all points, so that problems at the boundary will
immediately affect all points. Thus they are inherently more
sensitive to problematic boundary conditions. However, it is possible
that the boundary driven instabilities we observe with SGRID
could also affect finite differencing codes in some situations,
possibly at later times.

\begin{acknowledgments}

It is a pleasure to thank Peter Diener and David Hilditch
for useful discussions about BSSN gauge modes near black holes
and boundary conditions.
This work was supported by NSF grants PHY-0652874
and PHY-0855315.

\end{acknowledgments}




\bibliography{references}

\end{document}